\documentclass[twocolumn,preprintnumbers,amsmath,amssymb]{revtex4}
\usepackage{amsmath}
\usepackage{extarrows}
\usepackage{graphicx}
\usepackage{dcolumn}
\usepackage{bm}
\usepackage{amsmath}
\usepackage{multirow}
\usepackage{mathrsfs}
\usepackage{bbm}
\usepackage{euscript}
\usepackage{amssymb}
\usepackage{extarrows}
\usepackage{bbm}
\usepackage{graphicx}
\usepackage{epstopdf}
\usepackage{dcolumn}
\usepackage{bm}
\usepackage[all]{xy}
\usepackage{indentfirst}
\usepackage{amsmath}
\usepackage{multirow}
\usepackage{mathrsfs}
\usepackage{euscript}
\usepackage{amssymb}
\usepackage{appendix}
\usepackage{color,xcolor}
\usepackage{diagbox}
\newtheorem{theorem}{Theorem}

\newtheorem{lemma}{Lemma}
\newtheorem{corollary}{Corollary}

\begin{document}

\title{Entanglement polygon inequality in qudit systems}

\author{Xue Yang, Yan-Han Yang, Ming-Xing Luo}

\affiliation{The School of Information Science and Technology, Southwest Jiaotong University, Chengdu 610031, China}

\begin{abstract}
Entanglement is one of important resources for quantum communication tasks. Most of results are focused on qubit entanglement. Our goal in this work is to characterize the multipartite high-dimensional entanglement. We firstly derive an entanglement polygon inequality for the $q$-concurrence, which manifests the relationship among all the ``one-to-group'' marginal entanglements in any multipartite qudit system. This implies lower and upper bounds for the marginal entanglement of any three-qudit system. We further extend to general entanglement distribution inequalities for high-dimensional entanglement in terms of the unified-$(r, s)$ entropy entanglement including Tsallis entropy, R\'{e}nyi entropy, and von Neumann entropy entanglement as special cases. These results provide new insights into characterizing bipartite high-dimensional entanglement in quantum information processing.
\end{abstract}

\maketitle

\section{Introduction}

Entanglement as one of the most puzzling features in quantum mechanics reveals the fundamental insights into quantum correlations. This can be quantified by using special quantities such as the linear entropy for pure entanglement \cite{EOF1996,Hill1997,Wootters1998}. So far, various entanglement measures have been proposed for quantifying bipartite entangled systems \cite{Rungta2001,Kim2010T,Kim2010R,KimBarry2011,Khan2019}. The study of entanglement has inspired great applications in quantum communications \cite{Ekert1991,Horodecki2009}, the features of many-body systems \cite{Eisert2010,Amico2008}, and the potential limitations for quantum-enhanced technologies \cite{Dowling2003}.

The bipartite entanglement measure can be extended for multipartite entanglement. Interestingly, varied entanglement measures may be recovered from different bipartition of multipartite systems. Consider a tripartite entangled state $\rho_{123}$ as an example. There are six different bi-partite entanglements ${\cal E}_{1|23}$, ${\cal E}_{2|13}$, ${\cal E}_{3|12}$, ${\cal E}_{1|2}$, ${\cal E}_{1|3}$ and ${\cal E}_{2|3}$. Herein, ${\cal E}$ is an arbitrary bipartite entanglement measure, ${\cal E}_{i|jk}$ called as ``one-to-group'' marginal entanglement denotes the entanglement between a single partite $i$ and a group of partite $\{j,k\}$, and ${\cal E}_{i|j}$ is called ``one-to-one'' entanglement between the $i$-th and $j$-th particles. According to the spirit of the Coffman-Kundu-Wootters (CKW) inequality \cite{VCoffman}, the entanglement measure for a three-qubit system shows the following relationship as
\begin{eqnarray}
{\cal E}_{1|23}\geq {\cal E}_{1|2}+{\cal E}_{1|3}.
\label{eqn0}
\end{eqnarray}
This means the summation of one-to-one entanglements is upper bounded by the one-to-group  marginal entanglement. Unfortunately, the so-called entanglement monogamy inequality \cite{VCoffman,Terhal2004} is invalid for general entanglement measures. It might hold for special entangled systems. Especially, Osborne and Verstraete proved it for an arbitrary $n$-qubit systems \cite{TJOsborne} with the squared concurrence. Similar monogamy inequalities has been proved for multiple qubit states with different entanglement measures such as the squared negativity \cite{Kim2009N}, the squared entanglement of formation (EOF) \cite{Oliveira2014,Bai3,Bai2014}, the squared Tsallis-$q$ entropy \cite{Luo2016}, the squared R\'{e}nyi-$q$ entropy \cite{R2015}, and the squared unified-$(q, s)$ entropy \cite{Khan2019}. This is further improved as a class of tight $\alpha$-th power monogamy relations for multiqubit systems \cite{Fei1,Luo2015,Luo2016}.

Different from the CKW monogamy inequality, there is another polygon inequality which gives a upper bound for quantum marginal entanglements as \cite{Qian2018}:
\begin{eqnarray}
{\cal E}_{1|23}\leq {\cal E}_{2|13}+{\cal E}_{3|12}
\label{eqn00}
\end{eqnarray}
for any tripartite entangled pure state. As its stated in ref. \cite{Walter2013}, the collection of eigenvalues of the reduced density matrices of a pure entanglement form a convex polytope. This results in a local criterion for witnessing global multi-particle entanglement. The polygon inequalities are proved to be valid for arbitrary qubit pure states with respect to generic entanglement measures \cite{Qian2018}. Interestingly, high-dimensional entanglement opens intriguing perspectives in quantum information science \cite{Friis2019}, quantum communications \cite{Cerf2002,Sheridan2010,Mafu2013,Mirhosseini2015,Islam2017,Cozzolino2019} or technological advances \cite{Erhard2020}. Hence, a natural problem is to feature high-dimensional entangled systems. Inspired by recent results \cite{Qian2018,Walter2013}, we explore the polygon relationship among all the one-to-group marginal entanglement measures of pure entangled states. This can be regarded as resource sharing rules in distributed quantum applications.

The outline of the rest is as follows. In Sec.II, we present an entanglement polygon inequality in terms of the $q$-concurrence for any pure multipartite high-dimensional entanglement. This implies a lower bound for the one-to-group marginal entanglement of three-qudit system. In Sec.III, we establish entanglement distribution inequalities of multi-qudit entanglement based on the two-parametric unifed-$(r, s)$ entanglement. In Sec.IV, we construct a set of multipartite entanglement indicators. These results are further applied for featuring general high-dimensional entangled states and quantum networks. The last section concludes this paper.

\section{Entanglement polygon inequalities in terms of the $q$-concurrence}

In this section, our goal is to prove the entanglement polygon inequalities for multipartite high-dimensional entangled systems in terms of the $q$-concurrence.

\subsection{The $q$-concurrence}

In this subsection, we introduce a bipartite entanglement measure based on the parameterized entropy. A parameterized entropy function $F_{q}(\rho)$ of a quantum state $\rho$ on $d$-dimensional Hilbert space is recently defined \cite{Yang2021} as
\begin{eqnarray}
F_{q}(\rho)=1-{\rm{Tr}}\rho^q
\label{eqnFq0}
\end{eqnarray}
for each $q\geq2$, where $F_{q}(\rho)=0$ for any pure state $\rho$ while its maximal value is 1 for a completely mixed state $\frac{1}{d}\mathbbm{1}$ with the identity matrix $\mathbbm{1}$.

For any bipartite pure state $|\phi\rangle_{AB}$ on finite-dimensional Hilbert space ${\cal H}_A\otimes {\cal H}_B$, the $q$-concurrence is defined as \cite{Yang2021}:
\begin{eqnarray}
{\cal C}_{q}(|\phi\rangle_{AB})=F_{q}(\rho_A),
\label{cqd0}
\end{eqnarray}
where $\rho_A={\rm{Tr}}_B(|\phi\rangle_{AB}\langle\phi|)$ is the reduced density matrix of the subsystem $A$, and $F_{q}(\rho_{A})$ is defined in Eq.(\ref{eqnFq0}). For a mixed state $\rho_{AB}$ on ${\cal H}_A\otimes {\cal H}_B$, the $q$-concurrence is defined by the convex extension of all the pure decompositions as
\begin{eqnarray}
{\cal C}_{q}(\rho_{AB})=\inf_{\{p_i,|\phi_i\rangle\}}
\sum_ip_i{\cal C}_{q}(|\phi_i\rangle_{AB}),
\label{cqd1}
\end{eqnarray}
where the infimum is taken over all the possible pure state decompositions of $\rho_{AB}$, that is, $\rho_{AB}=\sum_ip_i|\phi_i\rangle_{AB}\langle\phi_i|$, $\{p_i\}$ is a probability distribution with $p_i\geq 0$ and $\sum_ip_i=1$. As its mentioned in ref. \cite{Yang2021}, the $q$-concurrence for a large $q$ may be applicable for featuring long-range entangled systems beyond the squared concurrence \cite{Hill1997}.

\textbf{Proposition 1}. The $q$-concurrence is an entanglement measure.

It has been proved that the $q$-concurrence is an entanglement monotone \cite{Wei}. Actually, according to the result in ref. \cite{Vidal2000}, the concavity of the entropy $F_{q}$ of an entanglement monotone ensures the monotonicity under local operations and classical communication (LOCC). The monotonicity of mixed states can be naturally inherited from the monotonicity of pure states via the convex roof extension. For the later use, we give following Lemma.

\begin{lemma} \cite{Yang2021} \label{additivity}
For a general bipartite entangled state $\rho_{AB}$ on Hilbert space ${\cal H}_{A}\otimes\mathcal{H}_{B}$, $F_{q}(\rho_{AB})$ satisfies the inequalities:
\begin{eqnarray}
|F_{q}(\rho_A)-F_{q}(\rho_B)|\leq F_{q}(\rho_{AB})\leq F_{q}(\rho_A)+F_{q}(\rho_B),
\label{eqnFq}
\end{eqnarray}
\end{lemma}
where $\rho_A$ and $\rho_B$ are reduced density matrices of the subsystem $A$ and $B$, respectively.

\subsection{Entanglement polygon inequalities with one-to-group entanglement}

Note that an $n$-qudit pure state on Hilbert space $\otimes_{i=1}^n{\cal H}_{i}$ can be generally represented by
\begin{eqnarray}
|\psi\rangle_{1\cdots{}n}=\sum^{n}_{j=0}\sum^{d_j-1}_{s_j=0}\alpha_{s_1\cdots s_n}|s_1\cdots s_n\rangle,
\label{purestate1}
\end{eqnarray}
where $\dim({\cal H}_{i})=d_i$ with $d_i\geq 2$ for $i=1, \cdots, n$; $\alpha_{s_1\cdots s_n}$ are the coefficients satisfying the normalization condition of $\sum_{s_1\cdots s_n}|\alpha_{s_1\cdots s_n}|^2=1$.

Consider a one-to-group entanglement ${\cal E}^{j|\overline{j}}$ of the pure state (\ref{purestate1}), where the subscript $j$ refers to the system of $j$, and $\overline{j}$ denotes all the systems except for the system $j$, and ${\cal E}$ is an arbitrary bipartite entanglement measure. These bipartite entanglements are also called as quantum marginal entanglements \cite{Walter2013}. Based on the present $q$-concurrence \cite{Yang2021} we derive the relationship among all the one-to-group  marginal entanglement measures for an arbitrary $n$-qudit system as the following theorem.

\begin{theorem}\label{relation0}
For any $n$-qudit pure entangled state $|\psi\rangle$ on Hilbert space $\otimes_{i=1}^n{\cal H}_{i}$, the following inequality holds
\begin{eqnarray}
{\cal C}^{j|\overline{j}}_{q}(|\psi\rangle)\leq \sum_{k\neq j,\forall k}{\cal C}^{k|\overline{k}}_{q}(|\psi\rangle)
\label{marginal1}
\end{eqnarray}
\end{theorem}
for any $q\geq2$.

\emph{Proof.} From the definition (\ref{cqd0}) we obtain
\begin{eqnarray}
{\cal C}^{j|\overline{j}}_{q}(|\psi\rangle)
\nonumber&=&F_{q}(\rho_{j})
\\
\nonumber
&=&F_{q}(\rho_{\overline{j}})
\\
&\leq&
\sum_{k\neq j,\forall k}F_{q}(\rho_{k})
\label{marginal2}
\\
&=&\sum_{k\neq j,\forall k}{\cal C}^{k|\overline{k}}_{q}(|\psi\rangle).
\label{marginal3}
\end{eqnarray}
Here, the inequality (\ref{marginal2}) is followed by iteratively using the subadditivity condition in Eq.(\ref{eqnFq}), and $F_{q}(\rho_{j})$ is the entropy of the reduced density matrix $\rho_{j}$ of the subsystem $j$. The inequality (\ref{marginal3}) is obtained from the definition (\ref{cqd0}). This has proved the result. $\Box$

For an arbitrary three-qudit system on Hilbert space ${\cal H}_{1}\otimes {\cal H}_{2}\otimes {\cal H}_{3}$, we have three one-to-other marginal entanglements, namely, ${\cal C}_{q}^{1|23}$, ${\cal C}_{q}^{2|13}$, ${\cal C}_{q}^{3|12}$. Theorem 1 implies special relationships among these quantities as follows.

\begin{corollary}
For an arbitrary three-qudit pure entangled state $|\psi\rangle$ on Hilbert space ${\cal H}_{1}\otimes {\cal H}_{2}\otimes {\cal H}_{3}$, the $q$-concurrence ${\cal C}_{q}^{i|jk}$ satisfy the following triangle inequalities as
 \begin{eqnarray}
|{\cal C}_{q}^{j|ik}(|\psi\rangle)-{\cal C}_{q}^{k|ij}(|\psi\rangle)|
&\leq& {\cal C}_{q}^{i|jk}(|\psi\rangle)
\nonumber
\\
&\leq & {\cal C}_{q}^{j|ik}(|\psi\rangle)+{\cal C}_{q}^{k|ij}(|\psi\rangle).
\label{relation1}
\end{eqnarray}

\label{relation01}
\end{corollary}

The right hand side of the inequality (\ref{relation1}) is a direct consequence of Theorem \ref{relation0}. Moreover, from the definition (\ref{cqd0}) and the triangle inequality in Lemma \ref{additivity}, we obtain
\begin{eqnarray}
{\cal C}^{i|jk}_{q}(|\psi\rangle)
&=&F_{q}(\rho_{jk})
\\
\nonumber
&\geq&|F_{q}(\rho_{j})-F_{q}(\rho_{k})|
\\
&=&|{\cal C}^{j|ik}_{q}(|\psi\rangle)-{\cal C}^{k|ij}_{q}(|\psi\rangle)|
\label{triangle10}
\end{eqnarray}
for any $i\not=j, k$. This completes the proof. $\Box$

\begin{figure}
\begin{center}
\resizebox{130pt}{120pt}{\includegraphics{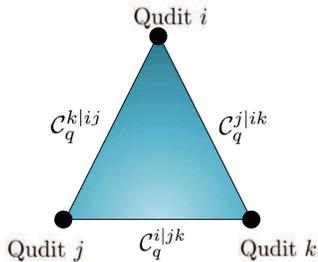}}
\end{center}
\caption{\small (Color online) Schematic entanglement distribution relationship of any tripartite entangled system. The length of each side represents correspondingly to the value of the marginal entanglement ${\cal C}^{j|ik}_{q}, {\cal C}^{i|jk}_{q}$ or ${\cal C}^{k|ij}_{q}$.}
\label{triangle}
\end{figure}

A simple geometric interpretation for the inequality (\ref{relation1}) is that ${\cal C}_{q}^{j|ik}(|\psi\rangle), {\cal C}_{q}^{k|ij}(|\psi\rangle)$ and ${\cal C}_{q}^{i|jk}(|\psi\rangle)$ consist of a triangle, that is, the one-to-group marginal entanglement $q$-concurrences can represent the lengths of the three edges of a triangle, as shown in Fig.\ref{triangle}. In fact, for a generic three-qubit system, the entanglement polygon inequality for the squared concurrence is firstly found as \cite{Zhu2015}:
\begin{eqnarray}
 {\cal C}_{2}^{i|jk}(|\psi\rangle)\leq {\cal C}_{2}^{j|ik}(|\psi\rangle)+{\cal C}_{2}^{k|ij}(|\psi\rangle).
\label{triangleC}
\end{eqnarray}
Obviously, the inequality (\ref{triangleC}) includes the result \cite{Zhu2015} as a special case for $q=2$. Moreover, the present result shows new features of long-range entanglement going beyond qubit systems with the squared concurrence \cite{Zhu2015}.

\subsection{Entanglement polygon inequalities with general bipartite entanglement}

In this subsection, we consider the relationship among the $q$-concurrence of any bipartition of multipartite entanglement. Consider an $m+n$-partite system in the state $|\psi\rangle$ on Hilbert space $\otimes_{i=1}^m{\cal H}_{A_i}\otimes_{j=1}^n{\cal H}_{B_j}$. Denote ${\bf{A}}=A_1\cdots A_m$ and ${\bf {B}}=B_1\cdots B_n$. Here, $|\psi\rangle^{\bf {A|B}}$ is defined by $|\psi\rangle^{\bf {A|B}}:=|\psi\rangle^{A_1\cdots A_m|B_1\cdots B_n}$, $A_i$'s and $B_j$'s are qudit subsystems. We have the following entanglement polygon inequalities for general bipartite entanglement.

\begin{theorem}
\label{bipartitionC}
For any entangled pure entangled state $|\psi\rangle_{\bf{AB}}$ the $q$-concurrence satisfies the following inequality
\begin{eqnarray}
{\cal C}_{q}^{\bf {A|B}}(|\psi\rangle)\leq \sum^m_{i=1}{\cal C}_{q}^{A_i|\overline{A}_i}(|\psi\rangle),
\label{eqntg2}
\end{eqnarray}
where ${\cal C}_{q}^{\bf {A|B}}$ denotes the $q$-concurrence under the bipartition $\bf {A|B}$. ${\cal C}_{q}^{A_i|\overline{A}_i}$ denotes the marginal entanglement between the subsystem $A_i$ and $\overline{A}_i$ denotes all the particles except $A_i$.

\end{theorem}

\emph{Proof.}  According to the $q$-concurrence (\ref{cqd0}), we get
\begin{eqnarray}
{\cal C}_{q}^{\bf {A|B}}(|\psi\rangle)
&=&F_{q}(\rho_{\bf {A}})
\nonumber
\\
&\leq&\sum^m _{i=1}F_{q}(\rho_{A_i})
\label{cqdmn0}
\\
&=&\sum^m_{i=1}{\cal C}_{q}^{A_i|\overline{A}_i}(|\psi\rangle).
\label{cqdmn}
\end{eqnarray}
Similar to the inequalities (\ref{marginal2}) and (\ref{marginal3}), the inequalities (\ref{cqdmn0}) and (\ref{cqdmn}) hold for any $m+n$-qudit pure state $|\psi\rangle_{\bf{AB}}$ on Hilbert space $\otimes_{i=1}^i{\cal H}_{A_i}\otimes_{j=1}^n\otimes {\cal H}_{B_{j}}$. This has completed the proof. $\square$

Inspired by ref. \cite{Qian2018}, define the distribution of a given amount of total entanglement, that is, ${\cal C}^{T}_{q}(|\psi\rangle)=\sum^n_{j=1}{\cal C}^{j|\overline{j}}_{q}(|\psi\rangle)$. It is easy to prove that
\begin{eqnarray}
{\cal C}^{j|\overline{j}}_{q}(|\psi\rangle)\leq \frac{{\cal C}^{T}_{q}}{2}(|\psi\rangle).
\label{aa}
\end{eqnarray}
By regarding $|\psi\rangle$ as an $n$-partite entanglement shared by $n$ parties $A_1, \cdots, A_n$, the present inequality (\ref{aa}) implies that for any single party $A_j$, its information (quantified by ${\cal C}^{j|\overline{j}}_{q}(|\psi\rangle)$) obtained from its reduced matrix $\rho_{A_j}$ is no more than half of the total. Theorem \ref{relation0} may provide an appropriate proposal in various quantum tasks. The more intriguing part is the present result shows new features of marginal entanglement distribution for high-dimensional quantum systems beyond the qubit entanglement \cite{Qian2018}. It also provides different inequalities beyond the linear polygon inequality \cite{Walter2013}.

Note Theorem 2 is not a special case of Theorem 1, they are not inequivalent. Take a 4-partite system as an example. In Theorem 2, we get new upper bounds of the marginal entanglement of $q$-concurrence on $4$-partite system under the partition 12 and 34, that is,
\begin{eqnarray}
{\cal C}^{12|34}_{q}\leq {\cal C}_{q}^{1|234}+{\cal C}_{q}^{2|134}.
\label{aa}
\end{eqnarray}
In Theorem 1, we provide the restrictions among all ``one-to-group'' entanglements between a single qudit $j\in\{1,2,3,4\}$ and the remaining ones in an arbitrary $n$-qudit system, for example,
\begin{eqnarray}
{\cal C}^{1|234}_{q}\leq {\cal C}_{q}^{2|134}+{\cal C}_{q}^{3|124}+{\cal C}_{q}^{4|123}.
\label{aa}
\end{eqnarray}
Obviously, Theorem 2 describes a global entangled distribution between the partial set $ij$ and the partial set $\{k,l\}$ while Theorem 1 describes the intrinsic entanglement between a single partite $i$ and the partial set $\{j,k,l\}$. That is, the new partition of ${\bf {A}}=A_1A_2\cdots A_m$ and ${\bf {B}}=B_1B_2\cdots B_n$ in Theorem 2 cannot be reformed into the partition of one qudit and all the other systems in Theorem 1.

\section{Entanglement polygon inequality in terms of unified-$(r, s)$ entropy entanglement}
\label{Uentropy}

In Sec.II, we have proved the polygon inequality for pure multipartite high-dimensional entanglement by using the entanglement measure of $q$-concurrence. In this section, we extend the result with the unified-$(r, s)$ entropy entanglement.

For a given state $\rho$ on Hilbert space $\mathcal{H}$, the unified-$(r, s)$ entropy \cite{Rathie1991,Hu2006,Rastegin2011} is defined as
\begin{eqnarray}
S_{r,s}(\rho)=\frac{1}{(1-r)s}[({\rm{Tr}}\rho^r)^s-1],
\label{U}
\end{eqnarray}
where $r, s\geq0$ and $r\neq1$, $s\neq0$. The unified-$(r, s)$ entropy converges to the R\'{e}nyi entropy \cite{Renyi1961,Hu2006,Rastegin2011},
\begin{eqnarray}
 \lim_{s \to0}S_{r,s}(\rho)=\frac{1}{1-r}\log_2 {\rm{Tr}} (\rho^r):=R_r(\rho).
\end{eqnarray}
The R\'{e}nyi entropy is not additive \cite{Linden2013}, but weak sub-additive \cite{Dam2002}. Especially, the R\'{e}nyi entropy satisfies the following inequality
\begin{eqnarray}
R_r(\rho_{A})-R_{0}(\rho_{B})\leq R_r(\rho_{AB})\leq R_r(\rho_{A})+R_{0}(\rho_{B})
\label{additivityR}
\end{eqnarray}
for any $r>0$ and $r\neq1$, where $\rho_{A}$ and $\rho_{B}$ are reduced density matrices of the subsystem $A$ and $B$, respectively. Moreover, the unified $(r, s)$-entropy tends to the Tsallis entropy \cite{Tsallis1988,Hu2006,Rastegin2011} as
\begin{eqnarray}
 \lim_{s \to1}S_{r,s}(\rho)=\frac{1}{1-r} ({\rm{Tr}} (\rho^r)-1):=T_r(\rho).
\end{eqnarray}
Here, the Tsallis entropy satisfies the subadditivity and triangle inequality as \cite{Audenaert2007}:
 \begin{eqnarray}
 |T_r(\rho_A)-T_r(\rho_B)|\leq T_r(\rho_{AB})\leq T_r(\rho_A)+T_r(\rho_B)
 \end{eqnarray}
for any $r>1$.

For the case of $r\to 1$, the unified $(r, s)$-entropy converges to the von Neumann entropy \cite{Nielsen,Hu2006,Rastegin2011} as
\begin{eqnarray}
 \lim_{r \to1}S_{r,s}(\rho)={\rm{Tr}}\rho\log_2 \rho:=S(\rho).
\end{eqnarray}
Note that the von Neumann entropy satisfies the following Araki-Lieb inequality \cite{AL} as
\begin{eqnarray}
 |S(\rho_A)-S(\rho_B)|\leq S(\rho_{AB})\leq S(\rho_A)+S(\rho_B),
\end{eqnarray}

For a given bipartite pure state $|\phi\rangle_{AB}$, the unified-$(r, s)$ entropy entanglement \cite{KimBarry2011} is given by
\begin{eqnarray}
{\cal U}_{r,s}(|\phi\rangle_{AB})=S_{r,s}(\rho_{A})
\label{U1}
\end{eqnarray}
for each $r, s\geq0$. For a given mixed state $\rho_{AB}$, its Unified-$(r, s)$ entropy entanglement via the convex-roof extension is defined by
\begin{eqnarray}
{\cal U}_{r,s}(\rho_{AB})=\inf_{\{p_i,|\phi_i\rangle\}}
\sum_ip_i{\cal U}_{r,s}(|\phi_i\rangle_{AB}),
\label{U2}
\end{eqnarray}
where the infimum is over all the possible pure state decompositions of $\rho_{AB}$.

For any $r\geq 1$ and $s\geq0$, the unified $(r, s)$-entropy satisfies the subadditivity \cite{Hu2006,Rastegin2011} as
\begin{eqnarray}
 S_{r,s}(\rho_{AB})\leq S_{r,s}(\rho_{A})+S_{r,s}(\rho_{B})
\label{U02}
\end{eqnarray}
and the Araki-Lieb inequality
\begin{eqnarray}
|S_{r,s}(\rho_{A})-S_{r,s}(\rho_{B})|\leq S_{r,s}(\rho_{AB}).
\label{U12}
\end{eqnarray}
Both inequalities are necessary to prove entanglement polygon inequalities.

\begin{theorem}
\label{relation4}
For any pure $n$-partite entanglement $|\psi\rangle$ on Hilbert space ${\cal H}_{A_1}\otimes \cdots\otimes {\cal H}_{A_n}$, the following inequality holds
\begin{eqnarray}
{\cal U}_{r,s}^{j|\overline{j}}(|\psi\rangle)\leq \sum^n_{k\neq j}{\cal U}_{r,s}^{k|\overline{k}}(|\psi\rangle),
\label{U11}
\end{eqnarray}
where ${\cal U}_{r,s}^{j|\overline{j}}(|\psi\rangle)$ represents the unified-$(r,s)$ entropy entanglement under the bipartition $j|\overline{j}$ for each $r\geq1$ and $s\geq0$, and $\overline{j}, \overline{k}$ is defined in Theorem \ref{relation0}.

\end{theorem}

\emph{Proof.}  From the definition (\ref{U1}), for any $r\geq1$ and $s\geq0$ we have
\begin{eqnarray}
{\cal U}_{r,s}^{j|\overline{j}}(|\psi\rangle)\nonumber&=&S_{r,s}(\rho_{j})
\\
\nonumber
&=&S_{r,s}(\rho_{\overline{j}})
\\
&\leq&
\sum_{k\neq j, \forall k}S_{r,s}(\rho_{k})
\label{margina40}
\\
&=&\sum_{k\neq j,\forall k}{\cal U}_{r,s}^{k|\overline{k}}(|\psi\rangle),
\label{margina41}
\end{eqnarray}
where $S_{r,s}(\rho_j)$ is the unified-$(r, s)$ entropy for the reduced density matrix $\rho_j$. Here, the inequality (\ref{margina40}) is obtained by iteratively using the subadditivity of the entropy $S_{r,s}(\rho)$ in the inequality (\ref{U02}). The inequality (\ref{margina41}) is followed from the definition (\ref{U1}). This has completed the proof. $\Box$

\begin{corollary}
\label{relation5}
For any pure tripartite entanglement $|\psi\rangle$ on Hilbert space ${\cal H}_1\otimes {\cal H}_2\otimes {\cal H}_3$, we obtain
\begin{eqnarray}
|{\cal U}_{r,s}^{j|ik}(|\psi\rangle)-{\cal U}_{r,s}^{k|ij}(|\psi\rangle)|&\leq &{\cal U}_{r,s}^{i|jk}(|\psi\rangle)
\nonumber\\
&\leq& {\cal U}_{r,s}^{j|ik}(|\psi\rangle)+{\cal U}_{r,s}^{k|ij}(|\psi\rangle)
\label{U03}
\end{eqnarray}
for any $r\geq1$ and $s\geq0$, where ${\cal U}_{r,s}^{i|jk}(|\psi\rangle)$ represents the unified-$(r, s)$ entropy entanglement with respect to the bipartition $i|jk$, $i\neq j \neq k\in\{1, 2,3\}$.

\end{corollary}

\emph{Proof.}  The right side of the inequality (\ref{U03}) is a directive consequence of Theorem \ref{relation4}. Similar to Corollary 1, according to the definition of the unified-$(r, s)$ entropy entanglement (\ref{U1}) and the triangle inequality (\ref{U12}), we obtain
\begin{eqnarray}
{\cal U}_{r,s}^{i|jk}(|\psi\rangle)
&=&S_{r,s}(\rho_{jk})
\\
\nonumber
&\geq&|S_{r,s}(\rho_{j})-S_{r,s}(\rho_{k})|
\\
&=&|{\cal U}_{r,s}^{j|ik}(|\psi\rangle)-{\cal U}_{r,s}^{k|ij}(|\psi\rangle)|
\label{triangleU}
\end{eqnarray}
for any $i\not=j, k$. This has completed the proof. $\Box$

\begin{theorem}\label{bipartion1}
For any pure $m+n$-partite entanglement $|\psi\rangle_{\bf{AB}}$ on Hilbert space $\otimes_{i=1}^m{\cal H}_{A_i}\otimes_{j=1}^n {\cal H}_{B_{j}}$, the unified-$(r, s)$ entropy entanglement satisfies the following inequality
\begin{eqnarray}
{\cal U}_{r,s}^{\bf {A|B}}(|\psi\rangle)\leq \sum^m_{i=1}{\cal U}_{r,s}^{A_i|\overline{A}_i}(|\psi\rangle)
\label{eqntg20}
\end{eqnarray}
for $r\geq1$ and $s\geq0$, where ${\cal U}_{r,s}^{\bf {A|B}}(|\psi\rangle)$ denotes the unified-$(r, s)$ entropy entanglement under the bipartition $\bf {A|B}$, and ${\cal U}_{r,s}^{A_i|\overline{A}_i}(|\psi\rangle)$ is the marginal entanglement of the subsystem ${A_i}$.
\end{theorem}

\emph{Proof.}  From the definition of the unified-$(r, s)$ entropy entanglement (\ref{U1}), we have
\begin{eqnarray}
{\cal U}_{r,s}^{\bf {A|B}}(|\psi\rangle)
&=&S_{r,s}(\rho_{\bf {A}})
\nonumber
\\
&\leq&\sum^m _{i=1}S_{r,s}(\rho_{A_i})
\label{tgU00}
\\
&=&\sum^m_{i=1}{\cal U}_{r,s}^{A_i|\overline{A}_i}(|\psi\rangle),
\label{tgU01}
\end{eqnarray}
where the inequality (\ref{tgU00}) is from the additivity of the unified-$(r, s)$ entropy (\ref{U02}). The equality (\ref{tgU01}) is obtained from the definition of the unified-$(r, s)$ entropy entanglement in Eq. (\ref{U1}). This has completed the proof. $\square$

From Theorem 3 a large class of polygon inequalities holds for multi-qudit systems in terms of the unified-$(r, s)$ entropy entanglement. That is, our result can reduce to every case of multi-qudit polygon inequalities such as R\'{e}nyi-$\alpha$ and Tsallis-$q$  entanglement and EOF for selective choices of $r$ and $s$. This reveals a generic frame for entanglement polygon equality concerning all well-known bipartite entanglement measures. Similar to Theorem 2. Theorem 4 also reveals the different entanglement distribution features going beyond Theorem 3.

Since the unified-$(r,s)$ entropy converges to the R\'{e}nyi entropy as the limiting case of $s$ tends to 0, we have
\begin{eqnarray}
 \lim_{s \to0}{\cal U}_{r,s}(\rho_{AB})={\cal R}_r(\rho_{AB}),
\end{eqnarray}
where ${\cal R}_r(\rho_{AB})$ is the R\'{e}nyi entropy entanglement of $\rho_{AB}$ \cite{Kim(2010)R}. From Corollary \ref{relation5} and inequality (\ref{additivityR}), we get the following corollary.

\begin{corollary}
\label{relation14}
For any pure tripartite entanglement $|\psi\rangle$  on Hilbert space ${\cal H}_1\otimes {\cal H}_2\otimes {\cal H}_3$ we have
\begin{eqnarray}
|{\cal R}_r^{j|ik}(|\psi\rangle)-{\cal R}_0^{k|ij}(|\psi\rangle)|&\leq& {\cal R}_r^{i|jk}(|\psi\rangle)
\nonumber
\\
&\leq& {\cal R}_r^{j|ik}(|\psi\rangle)+{\cal R}_0^{k|ij}(|\psi\rangle)
\label{}
\end{eqnarray}
for any $r\geq0$ with $r\neq1$, where ${\cal R}_r^{j|\overline{j}}$ denotes the R\'{e}nyi entropy entanglement under the bipartition $j|\overline{j}$.

\end{corollary}

Moreover, the unified-$(r,s)$ entropy \cite{KimBarry2011} converges to the Tsallis entropy as
\begin{eqnarray}
 \lim_{s \to1}{\cal U}_{r,s}(\rho_{AB})={\cal T}_r(\rho_{AB}),
\end{eqnarray}
where ${\cal T}_r(\rho_{AB})$ is the Tsallis entropy entanglement of $\rho_{AB}$ \cite{Kim2010T}. From Theorem 3, we get the following corollary.

\begin{corollary}
\label{relation04}

For any pure $n$-partite entanglement $|\psi\rangle$ on Hilbert space $\otimes_{i=1}^n{\cal H}_i$, we obtain
\begin{eqnarray}
{\cal T}_r^{j|\overline{j}}(|\psi\rangle)\leq\sum_{k\neq j,\forall k}{\cal T}_r^{k|\overline{k}}(|\psi\rangle)
\label{}
\end{eqnarray}
for any $r\geq1$, where ${\cal T}_r^{j|\overline{j}}$ denotes the Tsallis entropy entanglement under the bipartition $j|\overline{j}$. Especially, for any tripartite entanglement $|\psi\rangle$ on Hilbert space $\in {\cal H}_{1}\otimes {\cal H}_{2}\otimes {\cal H}_{3}$  we obtain
\begin{eqnarray}
|{\cal T}_r^{j|ik}(|\psi\rangle)-{\cal T}_r^{k|ij}(|\psi\rangle)|&\leq &{\cal T}_r^{i|jk}(|\psi\rangle)
 \nonumber
\\
&\leq&
{\cal T}_r^{j|ik}(|\psi\rangle)+{\cal T}_r^{k|ij}(|\psi\rangle)
\label{T03}
\end{eqnarray}
for each $r$ with $r>1$.

\end{corollary}

As $r$ tends to 1 we have
\begin{eqnarray}
 \lim_{r \to1}{\cal U}_{r,s}(\rho_{AB})={\cal E}_f(\rho_{AB}),
\end{eqnarray}
where ${\cal E}_f(\rho_{AB})$ is the EoF of $\rho_{AB}$  \cite{EOF1996,Hill1997,Wootters1998}. From Theorem 3, we get the following corollary.

\begin{corollary}\label{relation24}
For any pure $n$-partite entanglement $|\psi\rangle$ on Hilbert space ${\cal H}_{1}\otimes \cdots\otimes {\cal H}_{n}$ we obtain
\begin{eqnarray}
{\cal E}_f^{j|\overline{j}}(|\psi\rangle)\leq \sum_{k\neq j, \forall k}{\cal E}_f^{k|\overline{k}}(|\psi\rangle).
\label{}
\end{eqnarray}
Here, ${\cal E}_f^{j|\overline{j}}(|\psi\rangle)$ represent the EOF under the bipartition $j|\overline{j}$. Especially, for any tripartite entanglement $|\psi\rangle$ on Hilbert space ${\cal H}_{1}\otimes {\cal H}_{2}\otimes {\cal H}_{3}$  we obtain
\begin{eqnarray}
|{\cal E}_f^{j|ik}(|\psi\rangle)-{\cal E}_f^{k|ij}(|\psi\rangle)|&\leq& {\cal E}_f^{i|jk}(|\psi\rangle)
\nonumber
\\
&\leq& {\cal E}_f^{j|ik}(|\psi\rangle)+{\cal E}_f^{k|ij}(|\psi\rangle).
\label{T03}
\end{eqnarray}

\end{corollary}

To sum up, we find that general bipartite entanglement measures in terms of entanglement entropy satisfy polygon inequalities for high-dimensional systems. For arbitrary qubit pure states, Qian et al. \cite{Qian2018} prove a set of the polygon inequalities  with respect to entanglement measures including EOF \cite{EOF1996,Wootters1998}, concurrence \cite{Rungta2001}, negativity \cite{Vidal2002}. However, we still do not know whether the entanglement polygon inequality is valid for higher-dimensional systems apart from qubit systems in terms of the concurrence or negativity (see Table I), although their monogamy  properties for certain special high-dimensional systems have been investigated \cite{Kim2009N,Ou2007}.

\begin{table}[!htbp]\label{table}
\begin{center}
\setlength{\belowcaptionskip}{0.1cm}
\caption{A comparison of the entanglement polygon inequalities (EPI) in terms of various entanglement measures. Denote $\mathcal{C}_{q}$, $\mathcal{E}_f$, ${\cal T}_r$, ${\cal R}_r$, ${\cal U}_{r,s}$, $\mathcal{C}$, and $\mathcal{N}$ as the $q$-concurrence, entanglement of formation, Tsallis entropy entanglement and R\'{e}nyi entropy entanglement, unified-$(r, s)$ entropy entanglement, the concurrence, negativity, respectively.}
\begin{tabular}{|c|c|c|c|c|}
\hline
\diagbox{Measure}{Property}&EPI&System&Ref.\\ 
\hline
$\mathcal{C}_{q}$&$\surd$&$d^{\otimes n}(d\geq2)$&Ours\\
\hline
$\mathcal{E}_f$& $\surd$  &$2^{\otimes n}$&Ref.\cite{Qian2018}\\
\hline
$\mathcal{E}_f$& $\surd$ &$d^{\otimes n}(d\geq2)$&Ours\\
\hline
${\cal T}_r$& $\surd$($ q\geq1$)&$d^{\otimes n}(d\geq2)$&Ours\\
\hline
${\cal R}_r$& $\surd$($ r\geq0$)&$d^{\otimes 3}(d\geq2)$&Ours\\
\hline
${\cal U}_{r,s}$&$\surd$($ r\geq 1,s\geq0$)&$d^{\otimes n}(d\geq2)$&Ours\\
\hline
$\mathcal{C}$&$\surd$&$2^{\otimes n}$&Ref.\cite{Qian2018}\\
\hline
$\mathcal{C}$&$?$&$d^{\otimes n}(d\geq2)$&Ref.\cite{Qian2018}\\
\hline
$\mathcal{N}$&$\surd$&$2^{\otimes n}$&Ref.\cite{Qian2018}\\
\hline
$\mathcal{N}$ &?&$d^{\otimes n}(d\geq2)$&Ref.\cite{Qian2018}\\
\hline
\end{tabular}
\end{center}
\end{table}

\section{ A new type of entanglement indicator}

For an $n$-partite entanglement $|\psi\rangle$ on Hilbert space $\otimes_{i=1}^n\mathcal{H}_{i}$, define a class of the multipartite marginal entanglement indicator as
\begin{eqnarray}
\tau_{\cal E}(|\psi\rangle)=\min_{j}\{\sum^n_{k\neq j,k=1}{\cal E}^{k|\overline{k}}(|\psi\rangle)-{\cal E}^{j|\overline{j}}(|\psi\rangle)\},
\label{indicator}
\end{eqnarray}
where ${\cal E}$ is an arbitrary bipartite entanglement measure, and $\overline{j}$, $\overline{k}$ are given in Theorem \ref{relation0}.

Similarly, for an $n+m$-partite entanglement $|\psi\rangle$ on Hilbert space $\otimes_{i=1}^m\mathcal{H}_{A_i}\otimes_{j=1}^n\mathcal{H}_{B_j}$
\begin{eqnarray}
\hat{\tau}_{\cal E}(|\psi\rangle_{\bf {AB}})=\min_{i}\{\sum^{n+m}_{i=1}{\cal E}^{A_i|\overline{A}_i}(|\psi\rangle)-{\cal E}^{\bf {A|B}}(|\psi\rangle)\},
\label{indicator0}
\end{eqnarray}
where $\bf{A}$, $\bf{B}$, $A_i$ and $\overline{A}_i$ are defined in Theorem \ref{bipartitionC}.

Note that the EOF is defined as the von Neumann entropy of subsystem, and only one know the von Neumann entropy  satisfies the additivity \cite{Nielsen}:
\begin{eqnarray}
 S(\rho)=S(\varrho_A)+S(\varrho_B)
 \label{additivityS}
\end{eqnarray}
for any product state $\rho=\varrho_{A}\otimes\varrho_{B}$. $\varrho_A$ and $\varrho_B$ is the density matrix of the subsystem $A$, $B$, respectively.

Inspired by the additivity (\ref{additivityS}) and the idea in Ref.\cite{Bai3}, we obtain the following result.

\begin{theorem}\label{EOFgenuine}
For an arbitrary tripartite pure state $|\psi\rangle$ on Hilbert space $\mathcal{H}_A\otimes \mathcal{H}_B\otimes\mathcal{H}_C$, the multipartite entanglement indicator of EOF is zero if and only if $|\psi\rangle$ is product state, that is, $\tau_{\cal E}(|\psi\rangle)=0$ if and only if  $|\varphi\rangle_{ijk}=|\phi\rangle\otimes|\phi'\rangle$ for some pure states $|\phi\rangle$ and $|\phi'\rangle$, where $i \neq j\neq k\in \{A,B,C\}$.

\end{theorem}

\emph{Proof}. We first prove the necessity. Assume that the quantum state $|\psi\rangle$ on Hilbert space $\mathcal{H}_A\otimes \mathcal{H}_B\otimes\mathcal{H}_C$ is bipartite product state, that is, $|\psi\rangle$ has the following forms:
\begin{eqnarray}
|\varphi\rangle_{ijk}&=&|\phi\rangle_{i}\otimes|\phi\rangle_{jk},
\label{bipartite0}
\\
|\varphi\rangle_{ijk}&=&|\phi\rangle_{i}\otimes|\phi\rangle_{j}\otimes|\phi\rangle_{k},
\label{bipartite1}
\end{eqnarray}
where $|\phi\rangle_{i(j,k)}$ and $|\phi\rangle_{jk}$ are different pure states, and $i \neq j\neq k\in \{A,B,C\}$,

For the case (\ref{bipartite0}), from the definition of EoF it follows that ${\cal E}^{i|jk}_f=0$, ${\cal E}^{j|ik}_f={\cal E}^{jk}_f$, and ${\cal E}^{k|ij}_f={\cal E}^{jk}_f$. This implies that
\begin{eqnarray}
{\cal E}^{i|jk}_f={\cal E}^{j|ik}_f+{\cal E}^{k|ij}_f,
\label{bipartite2}
\end{eqnarray}
where ${\cal E}^{i|jk}_f$ denotes the bipartite entanglement under the bipartition $i$ and $jk$, ${\cal E}^{jk}_f$ is the entanglement between subsystems $i$ and $j$. This leads to $\tau_{{\cal E}_f}=0$. Similar result holds for the case (\ref{bipartite1}).

Now, we show the sufficiency. When the indicator $\tau_{{\cal E}_f}$ is zero, we get the equality (\ref{bipartite2}) for all $\{i,j\}\in \{A, B, C\}$. According to the definition of EoF \cite{Wootters1998}, for the left side of Eq.(\ref{bipartite2}) we have
 \begin{eqnarray}
 {\cal E}^{i|jk}_f(|\varphi)\nonumber&=&S(\varrho_{i})
 \\&=&S(\varrho_{jk})
 \label{Eof1}
\\&=&S(\varrho_{j})+S(\varrho_{k}),
 \label{Eof2}
 \end{eqnarray}
where $\varrho_{i}$ is the reduced density matrix $\rho_{i}$ of the subsystem $i$. Here,  the equality (\ref{Eof1}) is derived from the symmetry of von Neumann entropy \cite{Nielsen}. The equality (\ref{Eof2}) holds if and only if $\varrho_{jk}=\varrho_{j}\otimes \varrho_{k}$, that is, $|\varphi\rangle_{ijk}$ is biseparable state. This has completed the proof. $\Box$

\begin{figure}
\begin{center}
\resizebox{240pt}{180pt}{\includegraphics{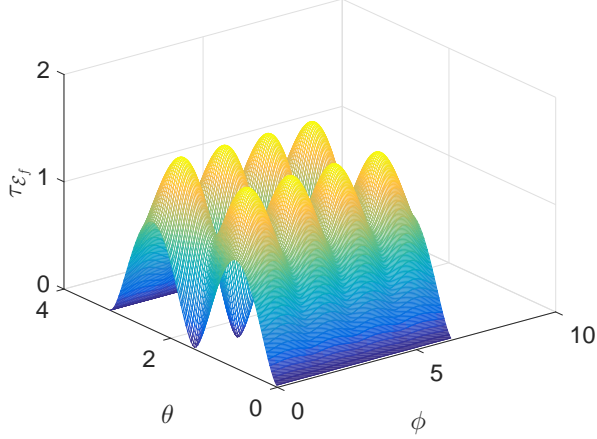}}
\end{center}
\caption{\small (Color online) The indicator $\tau_{{\cal E}_f}$ for a generalized three-qutrit GHZ state in Example 1.  }
\label{EPIGHZ3partite}
\end{figure}

\emph{Example 1}. Consider a generalized three-qudit GHZ state
\begin{eqnarray}
 |GHZ^g\rangle_{ABC}\nonumber&=&\sin\theta\cos\phi|000\rangle+\sin\theta\sin\phi|111\rangle
 \\&&+\cos\theta|222\rangle
\label{}
\end{eqnarray}
with $\theta\in [0, \pi]$ and  $\phi \in[0, 2\pi]$. According to the definition of EOF \cite{Wootters1998}, the values of three different bi-partite entanglements ${\cal E}^{i|jk}_f$s for all $i \neq j\neq k\in \{A,B,C\}$ can be calculated as
\begin{eqnarray}
{\cal E}^{i|jk}_f=-\sum^3_{i=1}\lambda_i\log_2\lambda_i,
\label{GHZEOF0}
\end{eqnarray}
where $\lambda_1=\sin^2\theta\cos^2\phi$, $\lambda_2=\sin^2\theta\sin^2\phi$, and $\lambda_3=\cos^2\theta$. From Eq.(\ref{indicator}), we have
 \begin{eqnarray}
\tau_{{\cal E}_f}(|GHZ^g\rangle)={\cal E}^{j|ik}_f+{\cal E}^{k|ij}-{\cal E}^{i|jk}_f.
\label{GHZEOF1}
\end{eqnarray}
Combing Eqs.(\ref{GHZEOF0}) and (\ref{GHZEOF1}), as shown in Fig.\ref{EPIGHZ3partite}, we state  that $\tau_{{\cal E}_f}(|GHZ^g\rangle)$ is nonnegative for $\theta\in [0, \pi]$ and  $\phi \in[0, 2\pi]$. $\tau_{{\cal E}_f}(|GHZ^g\rangle)=0$  when $|GHZ^g\rangle$ is separable, that is, corresponding to two cases of (i) $\theta=\pi$; (ii) $\theta=\frac{\pi}{2}$, $\phi=\frac{\pi}{2},\pi,\frac{3\pi}{2}$, or $2\pi$. For example, when $\theta=\frac{\pi}{2}$, $\phi=\frac{\pi}{2}$, or $\frac{3\pi}{2}$, the related state $|GHZ^g\rangle$ is changed into a separable state $|111\rangle$.

According to Theorem \ref{EOFgenuine}, the indicator is nonzero if a three-qudit pure state contains genuine tripartite entanglement. Howbeit, whether or not there exist similar results for the aforementioned bipartite entanglement measures is still unknown.

Next, we present some examples to show Theorems \ref{relation0}-\ref{bipartion1}.

\emph{Example 2}. Consider a three-qutrit W-class state \cite{Wstate},
\begin{eqnarray}
 |W_c\rangle_{123}=\frac{1}{\sqrt{6}}\sum^2_{i=1}(|i00\rangle+|0i0\rangle+|00i\rangle).
\label{}
\end{eqnarray}
According to the $q$-concurrence (\ref{cqd0}), we get
\begin{eqnarray}
{\cal C}_q^{i|jk}(|W_c\rangle)=1-\frac{2^q}{3^q}-\frac{2}{6^q}
\label{high002}
\end{eqnarray}
for each $i\neq j \neq k\in\{1, 2,3\}$. Thus, from  Eq.(\ref{indicator}) and  the symmetry of $|W_c\rangle$, we have
 \begin{eqnarray}
 \tau_{{\cal C}_q}(|W_c\rangle)={\cal C}_q^{i|jk}(|W_c\rangle)=-1+\frac{2^q}{3^q}+\frac{2}{6^q}.
 \label{W0}
\end{eqnarray}

Similarly, from Eqs.(\ref{U1}), (\ref{indicator}), and the symmetry of $|W_c\rangle$ we obtain
 \begin{eqnarray}
 \tau_{{\cal U}_{r,s}}(|W_c\rangle)&=&{\cal U}_{r,s}^{i|jk}(|W_c\rangle)
 \nonumber
 \\
 &=&\frac{1}{(1-r)s6^r}(4^r+2)^s-\frac{1}{(1-r)s}.
  \label{W1}
\end{eqnarray}
Note $\tau_{{\cal C}_q}(|W_c\rangle)$ and $\tau_{{\cal U}_{r,s}}(|W_c\rangle)$ are always positive. These implies Theorems  \ref{relation0} and  \ref{relation4} hold for a three-qutrit W-class state.

\emph{Example 3}. Consider an $m$-particle $d$-dimensional GHZ state \cite{GHZ}:
\begin{eqnarray}
 |GHZ\rangle_{12\cdots m}&=&\frac{1}{\sqrt{d}}\sum_{j=0}^{d-1}|j\rangle^{\otimes m}.
\label{}
\end{eqnarray}
According to Eqs.(\ref{cqd0}) and (\ref{U1}), from the symmetry of $|GHZ\rangle_{12\cdots m}$ we get
\begin{eqnarray}
&&{\cal C}^{j|\overline{j}}_{q}(|GHZ\rangle)=1-\frac{1}{d^{q-1}}
\label{GHZ1}
\\
&&{\cal U}^{j|\overline{j}}_{r,s}(|GHZ\rangle)=\frac{1-d^{rs-s}}{(1-r)sd^{rs-s}}
\label{GHZ2}
\end{eqnarray}
for any partite $j$ with $1\leq j\leq m$. These lead to the following equalities:
\begin{eqnarray}
&&\tau_{{\cal C}_q}(|GHZ\rangle)=(m-2){\cal C}^{j|\overline{j}}_{q}(|GHZ\rangle),
 \label{GHZ3}
 \\
&&\tau_{{\cal U}_{r,s}}(|GHZ\rangle)=(m-2){\cal U}^{j|\overline{j}}_{r,s}(|GHZ\rangle),
 \label{GHZ4}
\end{eqnarray}
where ${\cal C}^{j|\overline{j}}_{q}$ is defined in Eq.(\ref{GHZ1}) and ${\cal U}^{j|\overline{j}}_{r,s}$ is defined in Eq.(\ref{GHZ2}). It follows that $\tau_{{\cal C}_q}(|GHZ\rangle)$ and $\tau_{{\cal U}_{r,s}}(|GHZ\rangle)$ are nonnegative from the nonnegativity of ${\cal C}^{j|\overline{j}}_{q}(|GHZ\rangle)$ and ${\cal U}^{j|\overline{j}}_{r,s}(|GHZ\rangle)$.

\emph{Example 4}. Consider an $n$-partite complete graph-like network state $|G\rangle_{12\cdots n}$ \cite{Luo2021}, that is, any pair of two parties shares an EPR state, each party shares $n-1$ EPR states with others. There are $n(n-1)/2$ number of EPR states. Thus, from  Eqs.(\ref{cqd0}) and (\ref{U1}) we have
\begin{eqnarray}
&&{\cal C}^{j|\overline{j}}_{q}(|G\rangle)=\frac{(n-1)(2^{q-1}-1)}{2^{q-1}}
\\
&&{\cal U}^{j|\overline{j}}_{r,s}(|G\rangle)=\frac{(n-1)(1-2^{(r-1)s})}{2^{(r-1)s}(1-r)s}
\end{eqnarray}
for each partite $j$ with $1\leq j\leq n$. Similar to Eqs.(\ref{GHZ3}) and (\ref{GHZ4}), from the symmetry of the complete graph-like network state, it is easy to show that Theorems \ref{relation0} and  \ref{relation4} hold for the $n$-partite complete graph-like network state with respect to the $q$-concurrence and unified-$(r, s)$ entropy entanglement.

\emph{Example 5}. Consider an $n$-partite entangled quantum network $\mathcal{N}_q(\cal{A},\xi)$ \cite{Luo2021} consisting of EPR states and $m$-particle $d$-dimensional GHZ states, where $\cal{A}$ denotes the parties $\textsf{A}_1, \cdots, \textsf{A}_n$, and $\xi$ denotes entangled states. Assume any two parties $\textsf{A}_i$ and $\textsf{A}_j$ share the states $\varrho^1_{ij}, \cdots, \varrho^s_{ij}$, $\sigma^1_{ij}, \cdots, \sigma^t_{ij}$, and $\delta^1_{ij}, \cdots, \delta^k_{ij}$, where $\varrho$ denotes the density matrix of the EPR state \cite{EPR}: $|\phi\rangle=\frac{1}{\sqrt{2}}(|00\rangle+|11\rangle)$, and $\sigma$ is the density matrix of $m$-particle $d$-dimensional GHZ state \cite{GHZ}: $\frac{1}{\sqrt{d}}\sum_{j=0}^{d-1}|j\rangle^{\otimes m}$ with any integer $m\geq 3$, and $\delta=\frac{1}{d}\sum_{j=0}^{d-1}(|jj\rangle\langle jj|)$ is the reduced density matrix of any two subsystems obtained by tracing out other subsystems in an $m$-particle $d$-dimensional GHZ state. Here, any two parties $\textsf{A}_{i}$ and $\textsf{A}_{j}$ share the quantum state $\xi_{ij}$ given by
\begin{eqnarray}
\xi_{ij}=\mathop{\otimes}^{s_j}_{s=1}\varrho^s_{ij} \mathop{\otimes}^{t_j}_{t=1} \sigma^t_{ij}\mathop{\otimes}^{k_j}_{k=1}\delta^k_{ij},
\label{General0}
\end{eqnarray}
where $s_j$, $t_j$, and $k_j$ denote the numbers of $\varrho$, $\sigma$, and $\delta$, respectively. The total state $\rho_{\textsf{A}_1\cdots\textsf{A}_n}$ of the $n$-partite system is given by
\begin{eqnarray}
\rho_{\textsf{A}_1\cdots\textsf{A}_n}=\mathop{\otimes}^{n}_{\substack{i,j=1,\\i\neq j}}\xi_{ij}.
\label{networkstate}
\end{eqnarray}
Here, the total system may be regarded as an $n$-partite entanglement in high-dimensional Hilbert space. From Lemma 1, the local $F_q$ entropy of each party satisfies the subadditivity \cite{Yang2021}. Hence, from Eq.(\ref{cqd0}) we get
\begin{eqnarray}
{\cal C}^{\textsf{A}_j|\overline{\textsf{A}}_j}_{q}(\rho_{\textsf{A}_1\cdots\textsf{A}_n})
\nonumber
&=&F_q(\rho_{\overline{\textsf{A}}_j})
\\
\nonumber
&\leq&\sum_{k\neq j,\forall k}F_{q}(\rho_{k})
\label{networkstate0}
\\&=&\sum_{k\neq j,\forall k}{\cal C}^{k|\overline{k}}_{q}(\rho_{\textsf{A}_1\cdots\textsf{A}_n}).
\end{eqnarray}
Analogously, from the unified $(r, s)$-entropy entanglement (\ref{U1}) and the additivity  (\ref{U02}) we obtain
\begin{eqnarray}
{\cal U}^{j|\overline{j}}_{r,s}(\rho_{\textsf{A}_1\cdots\textsf{A}_n})
\leq\sum_{k\neq j}{\cal U}^{k|\overline{k}}_{r,s}(\rho_{\textsf{A}_1\cdots\textsf{A}_n}).
\end{eqnarray}
These implies Theorems  \ref{relation0} and \ref{relation4} hold for entangled quantum networks. Specifically, we consider the following chain network for showing the detail.

\begin{figure}
\begin{center}
\resizebox{90pt}{100pt}{\includegraphics{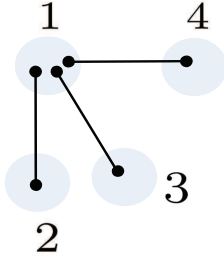}}
\end{center}
\caption{\small (Color online). A $4$-partite star network state consisting of three EPR states, where the subsystem 1 is regarded as an $8$-dimensional Hilbert space with local tensor decomposition.}
\label{star}
\end{figure}

\begin{figure}
\begin{center}
\resizebox{240pt}{190pt}{\includegraphics{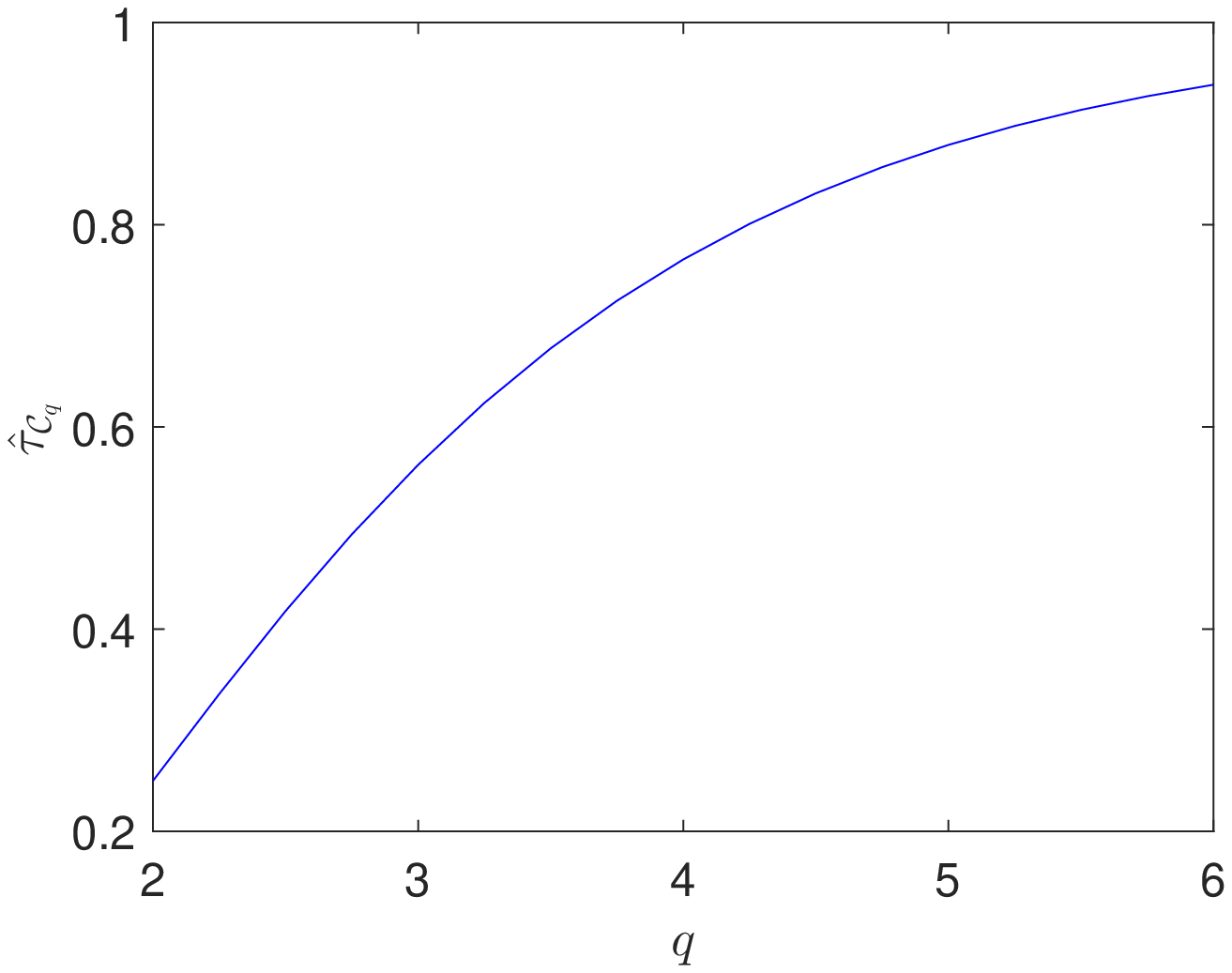}}

(a)

\resizebox{240pt}{190pt}{\includegraphics{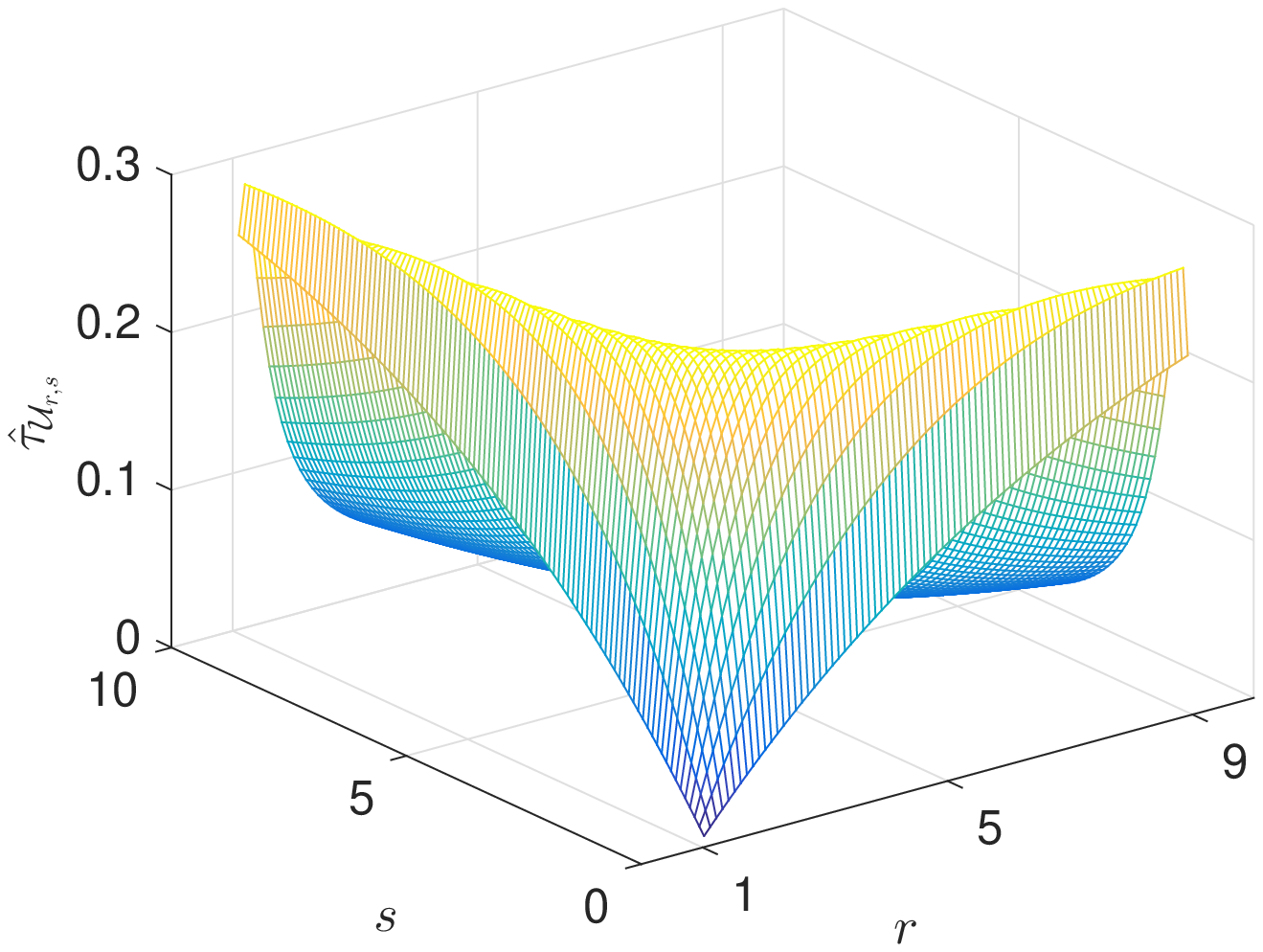}}

(b)
 \end{center}
\caption{\small (Color online) The indicator $\hat{\tau}_{{\cal C}_q}(|\varphi\rangle)$ and $\hat{\tau}_{{\cal U}_{r,s}}(|\varphi\rangle)$ in Example 6.  (a) The indicator $\hat{\tau}_{{\cal C}_q}(|\varphi\rangle)$ for $q\geq2$. (b) The indicator $\hat{\tau}_{{\cal U}_{r,s}}(|\varphi\rangle)$ for $1\leq r\leq 9$ and $0\leq s\leq10$. }

 \label{four}

\end{figure}

\emph{Example 6}. Consider a $4$-partite star quantum network consisting of three EPR states. It can be regarded as a $4$-partite pure state on $8\otimes 2 \otimes2 \otimes2$ dimensional Hilbert space $\otimes_{i=1}^4\mathcal{H}_i$, as illustrated in Fig.\ref{star}, where $\mathcal{H}_1$ is an $8$-dimensional space with basis $\{|0\rangle, \cdots, |7\rangle\}$ while $\mathcal{H}_i$ ($i\geq 2$) is a $2$-dimensional space with basis $\{|0\rangle, |1\rangle\}$.
Thus, the total state is given by
\begin{eqnarray}
|\varphi\rangle_{1234}
&=&\frac{1}{2\sqrt{2}}(|0000\rangle+|1001\rangle
+|2010\rangle+|3011\rangle
\nonumber\\
& &+|4100\rangle+|5101\rangle+|6110\rangle+|7111\rangle)_{1234}.
\label{high0}
\end{eqnarray}
The reduced state of the subsystems $1$ and $2$ is given by
\begin{eqnarray}
\rho_{12}&=&\frac{1}{8}(|00\rangle\langle00|+|41\rangle\langle41|
+|00\rangle\langle41|+|41\rangle\langle00|
\nonumber\\
&&+|10\rangle\langle10|+|51\rangle\langle51|+|10\rangle\langle51|+|51\rangle\langle10|
\nonumber\\
&&+|20\rangle\langle20|+|61\rangle\langle61|+|20\rangle\langle61|+|61\rangle\langle20|
\nonumber\\
&&+|30\rangle\langle30|+|71\rangle\langle71|+|30\rangle\langle71|+|71\rangle\langle30|
\label{re}
\end{eqnarray}
which has four nonnative spectrum $\frac{1}{4}$. Therefore, according to the $q$-concurrence in Eq.(\ref{cqd0}) we get
\begin{eqnarray}
&&{\cal C}_q^{12|34}(|\varphi\rangle)=\frac{4^{q-1}-1}{4^{q-1}},
\\
&&{\cal C}_q^{1|234}(|\varphi\rangle)=\frac{8^{q-1}-1}{8^{q-1}},
\\
&&{\cal C}_q^{2|134}(|\varphi\rangle)=\frac{2^{q-1}-1}{2^{q-1}}.
\label{high02}
\end{eqnarray}
Besides, we have
\begin{eqnarray}
&&{\cal C}_q^{34|12}(|\varphi\rangle)=\frac{4^{q-1}-1}{4^{q-1}},
\\
&&{\cal C}_q^{3|124}(|\varphi\rangle)={\cal C}_q^{4|123}(|\varphi\rangle)=\frac{2^{q-1}-1}{2^{q-1}}.
\label{high20}
\end{eqnarray}
Thus, taking the minimum of all possible bipartitions, the indicator $\hat{\tau}_{{\cal C}_q}$ from Eq.(\ref{indicator0}) is given by
\begin{eqnarray}
\hat{\tau}_{{\cal C}_q}(|\varphi\rangle)={\cal C}_q^{3|124}(|\varphi\rangle)+{\cal C}_q^{4|123}(|\varphi\rangle)-{\cal C}_q^{34|12}(|\varphi\rangle),
\end{eqnarray}
Moreover, from Eq.(\ref{U1}) we get that
\begin{eqnarray}
&&{\cal U}_{r,s}^{12|34}(|\varphi\rangle)=\frac{1-4^{rs-s}}{(1-r)s4^{rs-s}},
\\
&&{\cal U}_{r,s}^{1|234}(|\varphi\rangle)=\frac{1-8^{rs-s}}{(1-r)s8^{rs-s}},
\\
&&{\cal U}_{r,s}^{2|134}(|\varphi\rangle)=\frac{1-2^{rs-s}}{(1-r)s2^{rs-s}}.
\label{U04}
\end{eqnarray}
On the other hand, we also obtain
\begin{eqnarray}
&&{\cal U}_{r,s}^{34|12}(|\varphi\rangle)=\frac{1-4^{rs-s}}{(1-r)s4^{rs-s}},
\\
&&{\cal U}_{r,s}^{3|124}(|\varphi\rangle)=\frac{1-2^{rs-s}}{(1-r)s2^{rs-s}},
\\
&&{\cal U}_{r,s}^{4|123}(|\varphi\rangle)=\frac{1-2^{rs-s}}{(1-r)s2^{rs-s}}.
\label{U40}
\end{eqnarray}
Hence, taking the minimum of all possible bipartitions, we get
 \begin{eqnarray}
 \hat{\tau}_{{\cal U}_{r,s}}(|\varphi\rangle)={\cal U}_{r,s}^{{3|124}}(|\varphi\rangle)+{\cal U}_{r,s}^{4|123}(|\varphi\rangle)-{\cal U}_{r,s}^{12|34}(|\varphi\rangle).
\end{eqnarray}
Analogously, we illustrate the indicator $\hat{\tau}_{{\cal C}_q}(|\varphi\rangle)$ and  $\hat{\tau}_{{\cal U}_{r,s}}(|\varphi\rangle)$ in Fig.\ref{four}. Theorems \ref{bipartitionC} and \ref{bipartion1} hold for star network state in Eq.(\ref{high0}).

\section{Conclusion}

The distribution of multipartite entanglement is an important problem in entanglement theory and quantum information processing. Entanglement polygon inequality provides a distribution relationship of all the one-to-group marginal entanglements for arbitrary pure multi-partite entanglement. Based on the entanglement polygon inequality in qubit systems, a new genuine tripartite entanglement measure called the concurrence is recently proposed in ref.\cite{Xie2021}. It turned out that for any three-qubit pure state $|\psi\rangle_{123}$, the quantities of ${\cal C}^2_{1|23}(|\psi\rangle)$, ${\cal C}^2_{2|13}(|\psi\rangle)$ and ${\cal C}^2_{3|12}(|\psi\rangle)$ make up a triangle in geometry. In this case, the area of this squared concurrence triangle is useful for quantifying the genuine tripartite entanglement for tripartite qubit systems. A natural problem is whether the method is feasible for any higher dimensional systems by virtue of the inequality (\ref{relation1}). Besides, quantum coherence is also an essential ingredient in quantum information processing. Whether or not the coherence can be quantified by $q$-concurrence is valuable for consideration. Based on constraint relations among three bipartite coherences, a genuine tripartite coherence measure has been defined  \cite{Qian2020}. Similar to the quantification of genuine multipartite entanglement, a natural question arises: How to define a genuine multipartite coherence measure with respect to  all possible bipartite coherences.

To sum up, we have proved the entanglement polygon inequalities in terms of all the one-to-group $q$-concurrences between a single qudit and the others in any $n$-particle qudit system. Further, we showed that for a three-qudit system each entanglement polygon inequality provides an upper bound for a corresponding one-to-group marginal entanglement, while the triangle relation provides its lower bound. Additionally,  similar properties are proved for a two-parameter class of bipartite entanglement measures in terms of the unified-$(r, s)$ entropy entanglement. These results may intrigue new features of high-dimensional entanglement going beyond qubit systems.

\section*{Acknowledgments}

This work was supported by the National Natural Science Foundation of China (No.62172341) and Fundamental Research Funds for the Central Universities (No.2018GF07).


\end{document}